\documentclass[useASM]{mn2e}
\input{psfig.sty}
\usepackage{graphicx}

\begin{document}
\title{On the Spectral Lags of the Short Gamma-Ray Bursts\footnote{Correspondence author: Enwei Liang (lew@physics.unlv.edu)}}

\date{2002 February 06}
\pubyear{????} \volume{????} \pagerange{2} \twocolumn
\author[Yi et al. ]
 {Tingfeng Yi$^{1}$, Enwei Liang$^{1,2}$, Yiping Qin $^{1,3}$, Ruijing Lu$^{1,3,4}$\\
        $^1$Physics Department, Guangxi University, Nanning 530004, China
  \\
         $^2$Physics Department, University of Nevada, Las Vegas, NV 89154\\
         $^3$National Astronomical Observatories/Yunnan Observatory, Chinese Academy of Sciences, Kunming 650011, China\\
        $^4$Graduate School of The Chinese Academy of Sciences }

\date{Accepted Jan. 18, 2006.
      }

\pagerange{\pageref{firstpage}--\pageref{lastpage}} \pubyear{2001}

\maketitle
\label{firstpage}

\begin{abstract}
We present a detail analysis on the spectral lags of the short gamma-ray bursts (GRBs)
and compare them with that of the long GRBs by using the CGRO/BATSE GRB Catalog. Our
sample includes 308 short GRBs and 1008 long GRBs. The light curves of these GRBs are in
64 ms time bin and they have at least one long and intense pulse, which satisfies $\delta
T\geq 0.512 $ seconds at $c=1\sigma$ and $c_{\max}\geq 6\sigma$, where $\delta T$ is the
pulse duration, $c$ is the photon counts, and $\sigma$ is the standard error of the
background. We calculate the cross correlation function (CCF) for the light curves in
25-55 keV and 110-300 keV bands and derive the spectral lag by fitting the CCF with a
Gaussian model. Our results are as follows. (1)The spectral lag distribution of the short
GRBs is significantly different from that of the long GRBs. Excluding the statistical
fluctuation effect, a proportion of $\sim 17\%$ of the short GRBs have a negative
spectral lag, i.e., the hard photons being lag behind the soft photons. We do not find
any peculiar features from their light curves to distinguish these bursts from those with
a positive spectral lag. We argue that a more physical mechanism dominated the hard lag
may be hid behind the morphological features of the light curves. This should be a great
challenge to the current GRB models. We notice that this proportion is consistent with
the proportion of short GRBs correlated with nearby galaxies newly discovered by Tanvir
et al., although it is unclear if these short GRBs are indeed associated with the sources
originated at low redshift. (2)While the spectral lags of the long GRBs are strongly
correlated with the pulse durations, they are not for the short GRBs. However, the ratios
of the spectral lag to the pulse duration for the short and long GRBs are normally
distribution at 0.023 and 0.046, respectively, with the sample width, indicating that the
curvature effect alone could not explain the difference of the spectral lags between the
two types of GRBs. The hydrodynamic timescales of the outflows and the radiative
processes at work in GRBs might also play an important role as suggested by Daigne and
Mochkovitch.
\end{abstract}


\section{Introduction}
The phenomenon of gamma-ray bursts (GRBs) is still a great puzzle in astrophysics,
although much progress has been made (see reviews by Zhang \& M\'ez\'aros 2004; Piran
2005). Both the light curves and the spectra among GRBs are enormously various. There is
competing evidence that multiple GRB populations exist. The large sample detected by
CGRO/BATSE was identified as two types of GRBs, long-soft and short-hard GRBs, separated
by a duration of 2 seconds (Kouveliotou et al. 1993). It is well established that the
long GRBs are associated with energetic core-collapse supernovae(e.g., Galama et al.
1998; MacFadyen \& Woosley 1999; Bloom et al. 1999; Hjorth et al. 2003; Thomsen et al.
2004). Although it is well known that the short GRBs are phenomenologically different
from the long ones, such as variability time scale (Liang et al. 2002a; Nakar \& Piran
2002), fluence (Liang et al. 2002b), pulse interval and number of pulses in a burst
(Nakar \& Piran 2002), hardness ratio (Kouveliotou et al. 1993; Qin et al. 2000, 2001;
Dong \& Qin 2005; Qin \& Dong 2005), and spectrum (Ghirlanda et al. 2004; Cui, Liang, \&
Lu 2005), the issue that whether or not their energetics and host galaxies are different
from the long GRBs remains unknown being due to poor localization and lack of afterglow
observations before Swift era. Some groups paid much effort to search for the afterglows
for the short GRBs, but no detections were made (Kehoe et al. 2001; Gorosabel et al.
2002; Hurley et al. 2002; Klotz, Bo\"{e}r \& Atteia 2003). Swift satellite (Gehrels et
al. 2004) now is ushering us in a new era in short GRB study, and rapid progress was made
in the first Swift operation year (Gehrels et al. 2005; Hjorth et al. 2005a, b; Berger et
al. 2005; Fox et al. 2005). The precise localization of short GRB 050509B by Swift
(Gehrels et al. 2005) led to the first redshift measurement and host galaxy connection
(Bloom et al. 2005) for this class of GRBs. Among four short GRBs with known-redshift so
far, GRBs 050509B (z=0.2248, Bloom et al. 2005), 050724 (z=0.257; Berger et al. 2005),
and 050813 (z=0.72; Prochaska et al. 2005) coincide with the early-type stellar
population with no or little current star formation, favoring mergers of compact object
binaries as the progenitors of the short GRBs. one case, GRB 050709 (z=0.16), occurs at a
late-type dwarf galaxy with a star formation rate exceeding 0.5 $M_{\odot}$.yr$^{-1}$
(Prochaska et al. 2005). These breakthroughes rapidly improve our understanding on the
nature the short GRBs.

In this paper we present a detail analysis on the spectral lags of short GRBs. The
observed spectral lag is a common feature in high energy astrophysics (e.g., Norris et
al. 2000; Zhang et al. 2002). It is found that soft lag, soft photons being lag behind
the hard photons, is dominated in long GRBs (Norris et al. 2000; Wu \& Fenimore 2000;
Chen et al. 2005; Norris et al. 2005). The lag is also strongly correlated with the
luminosity (Norris et al. 2000) and the jet break times in afterglow light curves
(Salmonson et al. 2002). We here present a study on the spectral lags of the short GRBs
and compare them with that of the long bursts by using CGRO/BATSE GRB Catalog. The data
and our GRB sample selection are described in \S 2. We present our analysis method in \S
3, and compare the spectral lags in Short and Long GRBs in \S 4. We find that the
spectral lags of the long GRBs are correlated with the pulse duration ($\delta T$), while
the spectral lags of the short GRBs are not. We study the relation between the spectral
lag and the pulse duration and the ratio between these two quantities in \S 5. It is
interesting that about one-third of the short GRBs have a negative lag (hard lag, hard
photons being lag behind the soft photons). We present a further analysis on these short
GRBs in \S 6. Discussion and conclusions are presented in \S 7 and \S 8, respectively.

\section{Data and Sample Selection}
The sensitive GRB survey made by CRGO/BATSE presents a large and homogenous\footnote{It
is recently suggested that the BATSE short GRB sample may have two components(Tanvir et
al. 2005).} samples for both long and short GRBs. The light curves of these bursts are in
four energy bands, i. e., 25-55 keV, 55-110 keV, 110-300 keV, and $>300$ keV. We use the
light curve data in a time bin of 64 ms, which are concatenations of three standard BATSE
data types, DISCLA, PREB, and
DISCSC\footnote{ftp://cossc.gsfc.nasa.gov/compton/data/batse/}. We take the DISCLA data
before GRB trigger as the background of each GRB. We fit the DISCLA data by a linear
function, and then subtract the background with this linear model. Our long and short GRB
sample include 1008 and 308 GRBs, respectively. The light curves of these GRBs have at
least one long and intense pulse, which satisfies $\delta T\geq 0.512 $ seconds at
$c=1\sigma$ and $c_{\max}\geq 6\sigma$, where $c$ is the photon counts and $\sigma$ is
the standard error of the background.

\section{Analysis Method}
The cross correlation function (CCF) has been widely used to measure the time lag of two
light curves in two energy bands (Link, Epstein, \& Priedhorsky 1993; Cheng et al. 1995;
Band et al. 1997; Norris et al. 2000; Li et al. 2004; Chen et al. 2005; Norris et al.
2005). Assuming that $\{v_1\}$ and $\{v_2\}$  are the light curves in two different
energy bands, with $N$ data points in a time bin of $\Delta t$, the CCF as a function of
$\tau=k\Delta t$ is defined as
\begin{equation}
CCF(\tau=k\Delta t; v_1,v_2)=\frac{\sum_{t} v_{1}(t) v_{2}(t+\tau)}{N
\sqrt{\sigma'_{v_1}\sigma'_{v_2}}},
\end{equation}
where
\begin{equation}
{\sigma'}_v^2=\frac{1}{N}\sum\limits_{i=1}^{N}v_i^2.
\end{equation}
The CCF as a function of $\tau$ for two light curves with the same profile but having a
lag of $\tau_0$ symmetrically peaks at $\tau_0$. The observed light curves are mixed with
noises. The CCF hence may not symmetrically peak at $\tau_0$. We do not simply read off
the lag from the peak of the CCF curve. In order to reduce the scattering caused by noise
, we use a Gaussian function to fit the CCF curve, and take the peak of the Gaussian as
the lag. The lag derived by this way should be more precise than that directly read off
from the CCF. It can be a fraction of the time bin. A positive $\tau_0$ indicates a soft
lag, corresponding to an earlier arrival time of higher energy gamma-ray photons than
that of lower energy photons.

\begin{figure*}
\centering
\includegraphics[width=5in,angle=0]{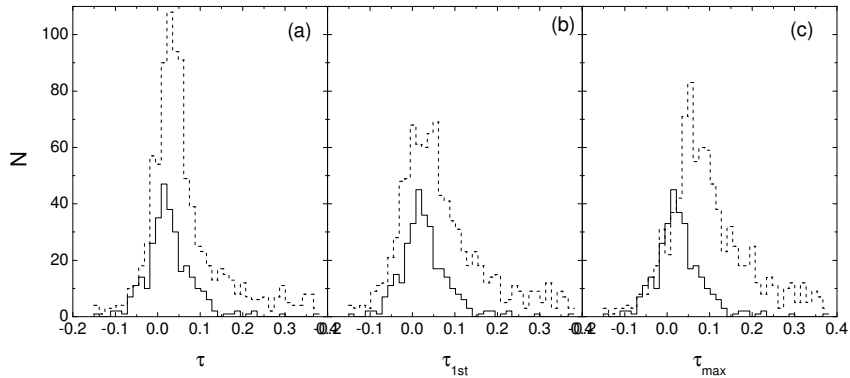} 
\caption{Comparisons of the spectral lags between the short GRBs (solid lines) and the
long GRBs: (a)the spectral lag derived from the whole time domain of a light curve;
(b)the spectral lag of the first pulse; (c) the maximum spectral lag among pulses in a
light curves. }\label{Fig1}
\end{figure*}

\section{Comparisons of spectral lags in Short and Long GRBs}
We calculate the lag between the light curves in 25-55 keV and 110-300 keV bands. The lag
derived by the CCF is an average time lag (${\tau}$) for the whole time domain. It is
found that some GRBs, such as GRB 960530 and GRB 980125, have significantly different
spectral lags in early and late epoches(Hakkila \& Giblin 2004). In principle, one could
not measure multiple lag components for two light curves by the CCF. Therefore, we also
calculate the spectral lag of each pulse in the light curves. The comparisons of the
distributions of the spectral lag measured from the whole light curves (${\tau}$),
spectral lag of the first pulse ($\tau_{1th}$), and the maximum spectral lag
\footnote{For a negative lag we take its minimum value as $\tau_{\max}$} among pulses
within a GRB ($\tau_{max}$) for the long and short GRBs are shown in Figure 1. We
summarize the features of these distributions in Table 1. The null hypothesis that these
distributions for the long and short GRBs are drawn from the same parent is tested by the
K-S test, and the results are also tabulated in Table 1. From Figure 1 and Table 1 one
can find that the spectral lags of the long GRBs are significantly larger than that of
the short GRBs. The most prominent difference is observed in the $\tau_{\max}$
distributions for the two kinds of GRBs. The mode of the $\tau_{\max}$ distribution of
the short GRBs is 0.015 seconds, while it is 0.057 seconds for the long GRBs. Soft lag is
dominated in long GRBs. About one-third of the short bursts, however, have a negative
spectral lag, i.e., the hard photons are lag behind the soft ones (hard lag). Further
discussion on these short GRBs is referred to \S 6.

\begin{table*}
\caption{Comparisons of the spectral lag and pulse duration distributions for the short
and long GRBs in our sample}
\tabcolsep0.08in
\begin{tabular}{lllllllll}

\hline \hline

&  &${\tau}$& &  $\tau_{1st}$  &  &   $\tau_{\max}$   &$\delta T^{*}$ \\
\hline
&short & long &short & long&short & long&short & long\\
\hline
Mean                       & 0.027   & 0.197  & 0.028 &     0.209  & 0.028  & 0.248&1.06&4.01\\
Median                     & 0.021   & 0.050  & 0.021 &     0.056  & 0.022  & 0.094&0.98&2.00\\
Mode                       & 0.015   & 0.023  & 0.015 &     0.024  &  0.015  & 0.057&0.85&1.25\\
Hard lag                  &30\%&17\%&31\%              &22\%&30\%&11\%\\
$P_{KS}$  &$2.8\times 10^{-14}$&&$2.7\times 10^{-17}$&&$ 1.4\times 10^{-43}$\\
\hline
\end{tabular}

$^{*}$ The $\delta T$ is corresponding to the pulse with $\tau_{\max}$ in a burst.

\end{table*}

\section{On the Relation between the Spectral Lag and the Pulse Duration}
As we shown above the spectral lags of the short GRBs are shorter than that of the long
GRBs. One could argue that as short GRBs are short, all timescales will scale with the
burst duration and the spectral lags will therefore be shorter. Liang et al. (2002a)
showed that the pulse duration of the short GRBs are indeed significantly shorter than
the long GRBs. We show the duration distributions of the pulses with $\tau_{\max}$ for
the short and long GRBs in our sample in Figure 2. Their means, medians, and modes are
also tabulated in Table 1. It is found that the pulse durations of the long GRBs are
significantly longer and have a larger dispersion than that of the short GRBs. One may
suspect that if there is a correlation between the spectral lag and the pulse duration.
Figure 3 shows the $\tau_{\max}$ as a function of the pulse duration for the short and
long GRBs. It is found that the two quantities are not correlated for the short GRBs, but
they are for the long GRBs. We measure this correlation by the Spearman correlation
analysis method. The correlation coefficient is 0.76 with a chance probability
$p<10^{-4}$ for the long GRBs.

The ratio of the spectral lag to the pulse duration may reflect some intrinsic properties
and present a clue to the mechanism dominated the spectral lag. We calculate this ratio
by $R=\tau_{\max}/\delta T$. Figure 4 shows the comparison of the $R$ distributions
between the long and the short GRBs. It is found that the $R$ distributions of the two
types of GRBs are narrowly clustered. We examine the normality of the two distributions
by the Shapiro-Wilk normality test. The normality test shows that the two distributions
are normal in a confidence level of $3\sigma$. We fit them by a normal distribution
function and derive $dN/dR\propto e^{-2\frac{(R-0.023)^2}{0.076^2}}$ and $dN/dR\propto
e^{-2\frac{(R-0.046)^2}{0.076^2}}$ for the short and long GRBs, respectively. The shapes
of the two distributions are almost the same but centering at 0.023 and 0.046 for the
short and long GRBs, respectively.

\begin{figure}
\centering
   \includegraphics[width=3in,angle=0]{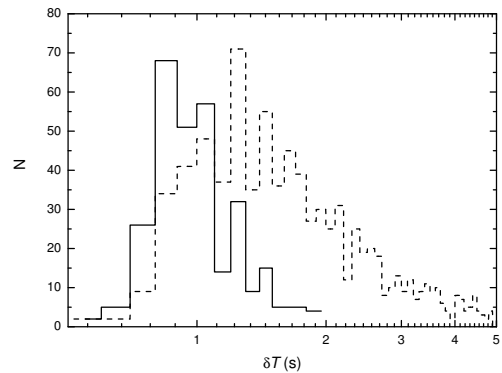} 
  \caption{Distributions of the the pulse durations for the short (Solid line) and long (dashed
  line) GRBs in our sample.}
  \label{Fig2}
\end{figure}

\begin{figure}
\centering
   \includegraphics[width=3in,angle=0]{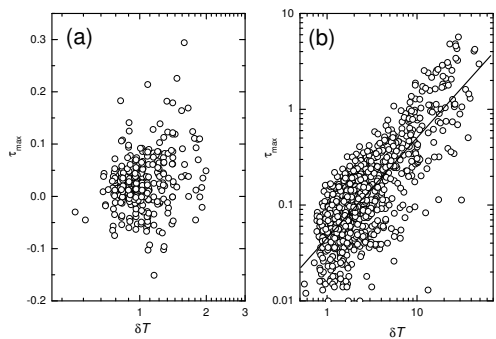} 
  \caption{Correlation between the spectral lag and pulse duration: (a)the short GRBs, (b) the long GRBs.
  The two quantities are not correlated for the short GRBs but they are for the long GRBs.
  The solid line in the right panel is the best fit to the two quantities for the long GRBs.}
  \label{Fig3}
\end{figure}

\begin{figure}
\centering
   \includegraphics[width=3in,angle=0]{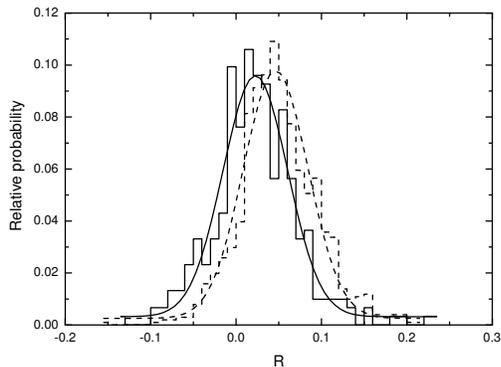} 
  \caption{Distributions of the ratio of $\tau_{\max}/\delta T$ for the short GRBs (solid step-line)
  and the long GRBs (dashed step-line). The solid and dashed lines are the fitting results by a Gaussian model for the
  short and long GRBs, respectively.}
  \label{Fig4}
\end{figure}

\section{The short GRBs with hard lag}
As we showed in \S 4, about one-third of the short GRBs have a hard lag. It is possible
that some of them are caused by statistical fluctuation when the soft lags of these GRBs
are very small. We thus consider only those short GRBs with $\tau_{\max}<-0.01$ seconds.
Figure 5 shows the distribution of these lags. We find that the statistical fluctuation
could not explain these hard lags. The distribution show a peak at $\sim -0.04$ seconds.
It is an incomplete normal distribution with a cutoff at -0.01, which can fitted by
$dN/d\tau_{\max}\propto e^{-2\frac{(\tau_{\max}+0.034)^2}{0.048^2}}$. This likely implies
that there are two components in the short GRB sample. The proportion of this component
calculated by this model is $\sim 17\%$. The fraction of the short GRBs with hard lags
caused by the statistical fluctuation is thus $\sim 13\%$, roughly consistent with the
fraction of the GRBs with a hard lag in the long GRB sample. Most recently, Tanvir et al.
(2005) find a position correlation between the short GRBs and the nearby galaxies and
propose a proportion of $5\%\sim 25\%$ of short-hard GRBs originated at low redshifts.
This proportion is roughly consistent with that of the short GRBs with a hard lag.
However, it is unclear if the short GRBs with a hard lag are indeed associated with the
sources originated at low redshift. This is very interesting as it is probably
challenging for many models. We thus present a further analysis on these bursts.

BATSE GRB trigger \# 2365 has the maximum hard lag among the short GRBs. The $T_{90}$ of
this burst is 1.536 seconds. Figure 6 show its light curves in 25-55 keV and $110-300$
keV bands. A prominent hard photon lag is observed. The lag is 0.151 seconds measured by
our method. The ratio of the rise time to the decay time of this pulse in 25-55 keV band
is 0.77 measured at the half of the amplitude of the light curve. This is consistent with
the statistical result from a sample of 92 short GRBs by Liang et al. (2002a). We further
compare the pulse duration distribution of the short GRBs having a soft lag with that
having a hard lag in Figure 7. We also could not find any peculiar features to
distinguish both of them.
\begin{figure}
\centering
\includegraphics[width=3in,angle=0]{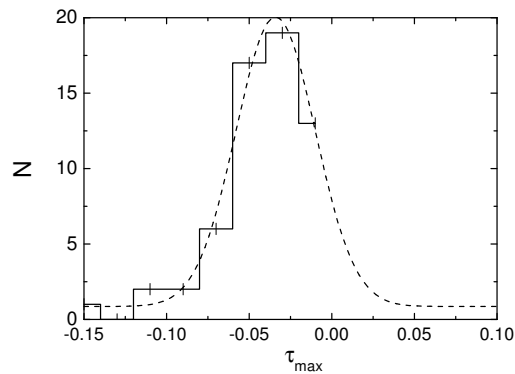} 
\caption{Distribution of the spectral lags for the short GRBs with $\tau_{\max}<-0.01$
seconds. The dashed line is the best fit by a Gaussian model.}\label{Fig5}
\end{figure}

\begin{figure}
\centering
\includegraphics[width=3in,angle=0]{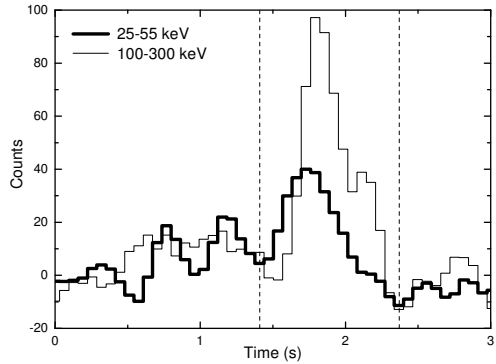} 
\caption{Light curves of BATSE GRB \# 2365 in 25-55 keV (thick line) and 110-300 keV
(thin line) bands. The two vertical lines mark the pulse with the most significant hard
lag in the short GRB sample.}\label{Fig6}
\end{figure}

\begin{figure}
\centering
\includegraphics[width=3in,angle=0]{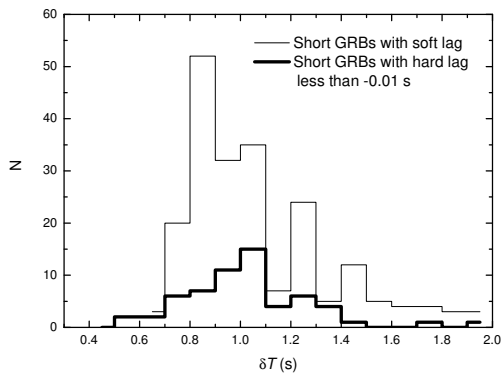} 
\caption{Comparison of the distributions of the pulse durations for the short GRBs with a
soft lag (thin line) and with a hard lag (thick line)}\label{Fig7}
\end{figure}

\section{Discussion}
The precision of the spectral lag measurement is determined by the time resolution and
the level of signal-to-noise of the light curves. In our analysis we derive the spectral
lag by fitting the CCF with a Gaussian model. This method reduces the statistical
fluctuation and the time bin selection effect. It gives a more precise measurement of the
spectral lag than that directly read off from the CCF, especially for light curves in a
short time period with a lag being comparable or smaller than the time resolution. We use
the time-tagged event (TTE) data to examine the reliability of our spectral lag
measurement. The TTE data are available only for the short GRBs. The TTE data record the
arrival time and energy band for each photon event. We construct the light curves of our
short GRB sample in 32 ms time bin and calculate their spectral lags. It is found that
both the spectral lags in 32 ms and 64 ms are comparable, suggesting our spectral lag
measurement is reliable.

The curvature effect has been extensively studies (e.g., Ryde \& Petrosian 2002; Qin
2002; Qin et al. 2004; Kocevski et al. 2003; Dermer 2004; Zhang et al. 2006; Dyks, Zhang,
\& Fan 2005), and it is suggested to responsible for the spectral lag (e.g., Salmonson
2000; Ioka \& Nakamura 2001; Shen, Song, \& Li 2005; Lu \& Qin 2005). Ryde (2005) found
that the spectral lag depends mainly on the pulse decay timescale. This is consistent
with the expectation of the curvature effect. In this scenario, the light curve in a
softer band tends to have a longer tail, and has a larger spectral lag relative to the
light curve in a harder band. We thus intuitively expect the ratio between the spectral
lag and the pulse duration is universal for different GRBs, if the decay phase of a pulse
is dominated by the curvature effect. As we shown in Figure 3, the distributions of the
ratios for the short and long GRBs are normal with the same width, suggesting the ratios
are universal within the long GRBs and short GRBs, respectively. However, they peak at
different values. The narrowly clustering of the distributions for the two types of GRBs
hints that the curvature effect may be partially responsible for the observed spectral
lag. The shift of the peaks of the two distributions, however, implies that this effect
alone could not account for the difference of the spectral lags between the short and
long GRBs. Other effects should play an important role on the pulse evolution, making the
difference of the spectral lags in the two types of GRBs. Several scenarios are also
involved to present an explanation for the spectral lag, such as an intrinsic cooling
effect of the radiating electrons (e.g. Zhang et al. 2002)  and a Compton reflection of a
medium (e.g., Ryde \& Petrosian 2002). Another possible scenario might be related to the
activity of the central engine and hydrodynamic timescale of the internal shocks (Daigne
\& Mochkovitch 1998; 2003). If the shell thickness is significantly smaller than the
initial shell separation, the hydrodynamic timescale may be relatively short enough and
the curvature effect governs the pulse evolution. If the shell thickness is comparable to
the initial shell separation or the outflows are continuous, the hydrodynamical timescale
may dominate the pulse evolution. Daigne \& Mochkovitch (1998; 2003) developed a model in
the framework of internal shock model (Rees \& M\'esz\'aros 1994) and found the observed
relation between luminosity and spectral lag (Norris et al. 2000) indeed can be
reproduced when the hydrodynamical timescale is taken into consideration.

The short GRBs with a hard lag is very interesting. This phenomenon is a challenge to
many models. The proportion of the short GRBs with negative lag is $\sim 17\%$, excluding
the statistical fluctuation effect. We notice a new finding by Tanvir et al. (2005), who
found a position correlation between the short GRBs and the nearby galaxies and proposed
a proportion of $5\sim 25\%$. This proportion is roughly consistent with that of the
short GRBs with a negative lag. However, it is unclear if the short GRBs with negative
lag are indeed associated with the sources originated at low redshift. Current GRB models
are hard to explain this phenomenon. We have tried to identify some distinguished
features of these bursts from those with a soft lag. We do not find such signatures from
their duration and the ratio of the rise time to the decay time. We guess that a more
physical mechanism dominated the hard lag may be hid behind the morphological features of
the light curves.

\section{Conclusions}
We have present a detail analysis on the spectral lags of the short GRBs and compare them
with that of the long GRBs by using the CGRO/BATSE GRB Catalog. Our sample includes 308
short GRBs and 1008 long GRBs. The light curves of these GRBs are in 64 ms time bin and
they have at least one long and intense pulse, which satisfies $\delta T\geq 0.512 $
seconds at $c=1\sigma$ and $c_{\max}\geq 6\sigma$. We calculate the CCFs for the light
curves in 25-55 keV and 110-300 keV bands and derive the spectral lags by fitting the CCF
with a Gaussian model. We summary our results as follows.

(1)The spectral lag distribution of the short GRBs is significantly different from that
of the long GRBs.  The null hypothesis that the spectral lag distributions of the two
kinds of GRBs are from the same parent is rejected, which is tested by the K-S test.

(2)Excluding the statistical fluctuation effect, a proportion of $\sim 17\%$ of the short
GRBs have a negative spectral lag. We do not find any peculiar features from their light
curves to distinguish these bursts from those with a positive spectral lag. We argue that
a more physical mechanism dominated the hard lag may be hid behind the morphological
features of the light curves. This should be a great challenge to the current GRB models.
We notice that this proportion is consistent with the proportion of short GRBs correlated
with nearby galaxies newly discovered by Tanvir et al., although it is unclear if these
short GRBs are indeed associated with the sources originated at low redshift.

(3)While the spectral lags of the long GRBs  are strongly correlated with the pulse
duration, the spectral lags of the short GRBs are not. However, the distributions of the
ratio $\tau_{max}/\delta T$ for two kinds of GRBs are normal in a confidence level of
$3\sigma$, and they have the same shapes but centering at 0.023 and 0.046 for the short
and long GRBs, respectively. This result indicates that the curvature effect alone could
not explain the difference of the spectral lags between two types of GRBs. The radiative
processes at work in GRBs and the hydrodynamic timescales of the outflows may also play
an important role as suggested by Daigne and Mochkovitch (1998, 2003).

We are grateful to the referee, Fr\'eder\'ic Daigne, for his valuable comments. We also
thank Bing Zhang (UNLV) for helpful discussion. Tingfeng Yi thanks Zhenqiang Tan (GXU)
and Xuefei Chen (YNAO) for their kind help. This work is supported by the National
Natural Science Foundation of China ( No.10463001 and No.10273019) and the Special Funds
for Major State Basic Research Projects (``973").

\end{document}